\documentclass[entropy,article,accept,pdftex,moreauthors]{Definitions/mdpi} 

\firstpage{1} 
\pubvolume{1}
\issuenum{1}
\articlenumber{0}
\pubyear{2024}
\copyrightyear{2024}
\externaleditor{Academic Editor: Firstname \linebreak Lastname}
\datereceived{25 July 2024} 
\daterevised{28 August 2024} 
\dateaccepted{3 September 2024 } 
\datepublished{ } 
\hreflink{https://doi.org/} 

\Title{Exploring the Diversity of Nuclear Density through Information~Entropy}

\TitleCitation{Exploring the Diversity of Nuclear Density through Information Entropy}

\Author{{Wei-Hu Ma} 
 $^{1,2}$ and {Yu-Gang} 
 Ma $^{1,2,}$*\orcidB{}}

\AuthorNames{Wei-Hu Ma and Yu-Gang Ma}

\AuthorCitation{{Ma, W.H.;} 
 Ma, Y.G.}

\address{%
$^{1}$ \quad Key Laboratory of Nuclear Physics and Ion-beam Application (MOE), Institute of Modern Physics, Fudan~University, Shanghai 200433, China; {maweihu@fudan.edu.cn} 
\\
$^{2}$ \quad Shanghai Research Center for Theoretical Nuclear Physics, NSFC and Fudan University, \mbox{Shanghai 200438, China}}

\corres{Correspondence: mayugang@fudan.edu.cn}

\abstract{This study explores the role of information entropy in understanding nuclear density distributions, including both stable configurations and non-traditional structures such as neutron halos and $\alpha$-clustering. By quantifying the uncertainty and disorder inherent in nucleon distributions in nuclear many-body systems, information entropy provides a macroscopic measure of the physical properties of the system. A more dispersed and disordered density distribution results in a higher value of information entropy. This intrinsic relationship between information entropy and system complexity allows us to quantify uncertainty and disorder in nuclear structures by analyzing various geometric parameters such as nuclear radius, diffuseness, neutron skin, and cluster structural features.}

\keyword{information entropy; nuclear density distribution;  halo nuclei;  nuclear clustering} 

\begin{document}


\section{Introduction}

Information entropy, a concept introduced by Claude Shannon to measure uncertainty in communication theory \cite{Shannon}, has extensive applications in many scientific disciplines. In nuclear physics, a pioneering study on information entropy was performed in Ref.~\cite{YGMa}, where the multiplicity information was proposed and applied to search for nuclear liquid gas phase transition \cite{Ma03b,Ma23}.  Applications of entropy or information entropy have been extended in different physics fields, e.g., Refs.~ \cite{entropy, Longair, Moustakidis, Massen, Csernai1, CWMa1, ZCao, CWMa2,  JXu, Csernai2, Lichtenberg, CWMa4, HLWei}. It has become a valuable tool for the study of complex systems, providing a way to quantify uncertainty and disorder in nuclear systems. Its applications range from the study of nuclear structure to the analysis of the dynamics of particle production in high-energy collisions \cite{Biales,FLi,Pu,DengXG}. As the field advances, the integration of information entropy with theoretical models and experimental data could lead to deeper insights into the fundamental properties of nuclear matter.

Research on weakly bound nuclei has revealed unique quantum phenomena such as halo and skin formation \cite{Tani,Xu,Bagchi,Fang,XuFR,MaZhang,Chen,ZhouL,SuY,YanTZ,ZhengZY,GengKP,AnR,HuoEB}, molecular orbitals, and cluster structures \cite{Oertzen,WMa,Horiuchi,HeWB,Zhou,WangYZ,WangSS, CaoYT,CaoRX,Ma_NT,ZhangYX}. These discoveries have reshaped our understanding of nucleon distributions in atomic nuclei, which often deviate from idealized spherical or geometric models. The resulting patterns are complex and irregular, driven by factors such as nucleon interactions, energy levels, and clustering dynamics.

These non-uniform nucleon distributions result in regions of varying density, from densely packed areas to sparser edges. This irregularity directly affects the strength and range of nuclear forces, which, in turn, affects how nuclei interact and behave during reactions \cite{Ozawa,Fang2}. This complexity challenges traditional models of nuclear structure and underscores the need for more nuanced investigations.

Understanding these dynamics is critical to accurately modeling nuclear structures and predicting their behavior in various situations. By studying these patterns, researchers can gain deeper insights into the coexistence of mean-field and cluster dynamics, shedding light on fundamental nuclear properties and processes. This knowledge is central to the advancement of nuclear physics and can guide future studies of nuclear interactions and~reactions.

In analogy to the information entropy for discrete values, the differential entropy for a continuous random variable $X$ with probability density function $f(x)$ is defined \cite{Thomas} as
\begin{equation}
\begin{aligned}
S(X) = -\int_{X}f(x) \ln f(x)dx,
\end{aligned}
\end{equation}
with the normalization condition $\int_{X}f(x)dx=1$. This approach supports the broader application of information entropy in nuclear physics, allowing the analysis of continuous~distributions. 

The purpose of this study is to explore how information entropy is used in nuclear physics to understand nuclear density distributions, including both stable configurations and unconventional arrangements such as the halo structure and the $\alpha$ clustering structure. The total information entropy of a system can serve as a comprehensive measure of physical information, providing a view of the nucleon distribution in a nuclear many-body system that reflects the disordered information of spatial configurations.


\section{Information Entropy of Diverse Nuclear Structures}
\subsection{Information Entropy of the Density Distribution with Woods--Saxon Type}

The nuclear Woods--Saxon distribution is a widely used model in nuclear physics to describe the spatial distribution of nucleons in an atomic nucleus \cite{WS, DATATables, WSdensity,Shou}. It is essential for the study of heavier nuclei with diffuse surfaces and provides a smooth transition from the high-density core to the low-density edge, accurately reflecting the structure of many nuclei. Its flexibility allows it to fit experimental data, making it a valuable tool in nuclear physics for analyzing structures and processes such as reactions, scattering, and decay.

The nuclei $^{16}$O, $^{40}$Ca, $^{116}$Sn, and $^{208}$Pb with Woods--Saxon type density distribution~\cite{WSdensity}
\begin{equation}
\begin{aligned}
\rho(r) = \frac{\rho_{0}}{1+e^{(r-R_{0})/a}}
\end{aligned}
\end{equation}
are considered as examples to examine the information entropy. The normalization coefficient $\rho_{0}$, the diffuseness parameter $a$, and the radius $R_{0}$ satisfy the normalization condition
\begin{equation}
\begin{aligned}
\int_{0}^{\infty}\frac{4\pi\rho(r)r^{2}}{W}dr=1
\end{aligned}
\end{equation}
where $W$ can be the number of protons $Z$, neutrons $N$, or nucleons $A=N+Z$. The parameters $R_{0}$ and $a$ describe the spatial distribution of protons and neutrons in a nucleus. The $R_{0}$ parameter for proton and neutron distributions is denoted as $R_{0p}$ and $R_{0n}$, respectively \cite{WSdensity}. They are given by $R_{0p}=1.81Z^{1/3}-1.12$ and $R_{0n}=1.49N^{1/3}-0.79$. The $a$ parameter for the proton and neutron distributions is defined as $a_{p}=0.47-0.00083Z$ and $a_{n}=0.47+0.00046N$, respectively. The diffuseness parameters $a_{p}$ and $a_{n}$ of the nuclei $^{16}$O, $^{40}$Ca, $^{116}$Sn, and $^{208}$Pb are scaled by the same parameter $n_{a}$ to serve as a variable. {The diffuseness parameters are modified by the scaling factor $n_{a}$, which serves as a variable to adjust the diffuseness parameters and, consequently, alter the density distribution. By using Equation (4), we can calculate the entropy as a function of $n_{a}$ and analyze the correlation between entropy and the diffuseness parameter.} 

\begin{figure}[H]
\includegraphics[width=0.650\hsize]{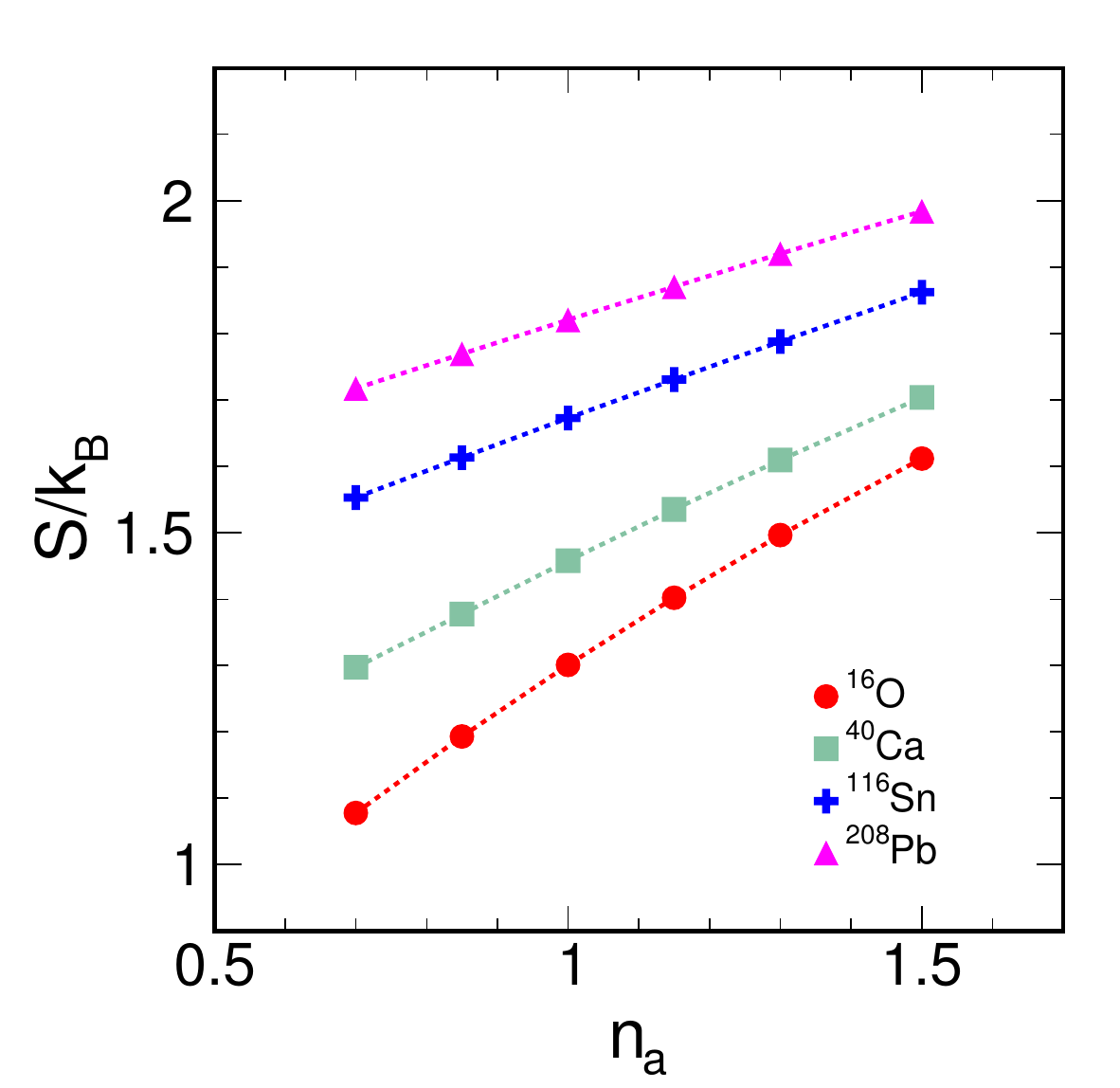}
\caption{(color online) Information entropy varying as the diffuseness scaling parameter $n_{a}$ for nuclei $^{16}$O, $^{40}$Ca, $^{116}$Sn, and $^{208}$Pb with Woods--Saxon type density distribution for {nucleons.} 
}
\label{fig1}
\end{figure}

The information entropy in the nuclear system with radial density distribution of the Woods--Saxon type is defined as
\begin{equation}
\begin{aligned}
S/k_{B} = -\int_{0}^{\infty}\frac{4\pi\rho(r)r^{2}}{W}\ln\frac{4\pi\rho(r)r^{2}}{W}dr,
\end{aligned}
\end{equation}
where $k_{B}$ is the Boltzmann constant, introduced to maintain the consistency of units in information entropy, rather than to highlight its thermodynamic significance. As shown in Figure \ref{fig1}, the information entropy tends to increase linearly with an increasing diffusion parameter, indicating an increase in the uncertainty of the radial distribution of nucleons. Heavier nuclei exhibit larger information entropy values, suggesting that a greater number of nucleons enhances the complexity of the nuclear distribution and increases the uncertainty in the nuclear radial distribution. However, the rate of increase in information entropy for heavier nuclei is less pronounced, indicating a potential saturation effect as the system reaches a certain level of complexity.

\begin{figure}[H]
\includegraphics[width=0.650\hsize]{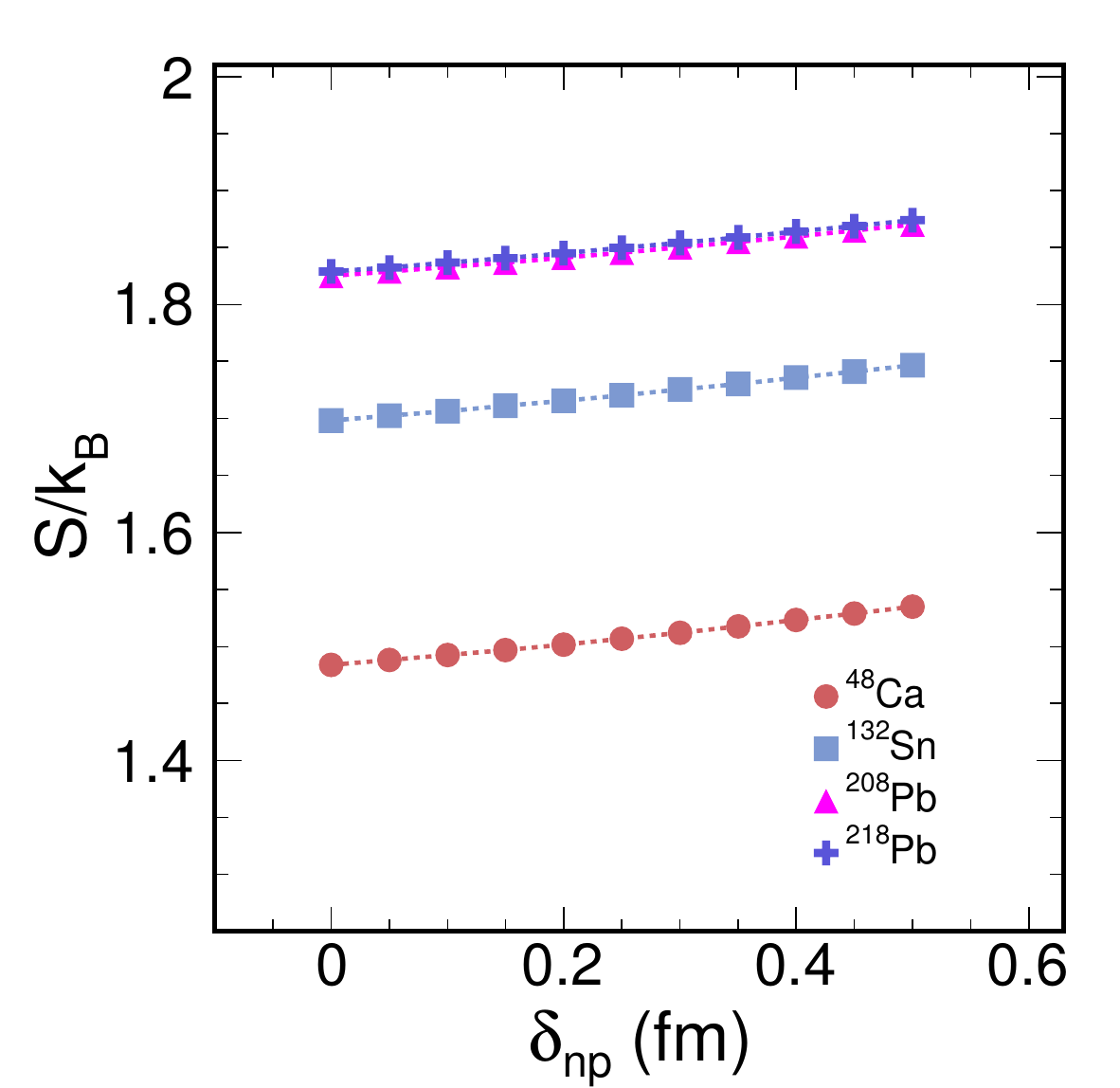}
\caption{(color online) Information entropy varying as neutron skin thickness for nuclei $^{48}$Ca, $^{132}$Sn, $^{208}$Pb, and $^{218}$Pb with Woods--Saxon type density distribution for {nucleons.}
}
\label{fig2}
\end{figure}

The information entropy is observed to be a function of the neutron skin thickness in neutron-rich nuclei. This thickness is defined as the difference between the square root radii of neutrons and protons, expressed as $\delta_{np}=R_{n}-R_{p}$, where $R_{n}$ and $R_{p}$ are the square root radii for neutrons and protons, respectively. By fixing the parameters $R_{p}$, $a_{n}$, and $a_{p}$ and varying $R_{n}$, we can examine the variable $\delta_{np}$. {With $R_{n}$ varying, the density distribution changes accordingly, allowing us to calculate the entropy corresponding to each value of $\delta_{np}$. Consequently, the information entropy can be examined as a function of neutron skin thickness in neutron-rich nuclei.} As shown in Figure \ref{fig2}, the relationship between information entropy and neutron skin thickness is linearly positive; as the neutron skin becomes thicker, the information entropy increases in a linear mode. Among the neutron-rich nuclei studied, such as $^{48}$Ca, $^{132}$Sn, $^{208}$Pb, and $^{218}$Pb, the heavier nuclei, such as $^{218}$Pb and $^{208}$Pb, have higher information entropy compared to $^{132}$Sn, followed by $^{48}$Ca. 
 Furthermore, all these nuclei show almost the same trend in the increase in the information entropy with respect to the neutron skin thickness. The neutron skin thickness parameter is crucial for gaining insight into the nuclear symmetry energy and the equation of state, which are fundamental aspects of nuclear physics. Information entropy provides new insights into the study of these phenomena.

\subsection{Information Entropy of The He/Li Isotope with Neutron-Rich Tail}

In loosely bound nuclei, the neutron density distribution extends into an unusually long tail, known as the neutron halo phenomenon. Despite the low density that characterizes the halo, its presence significantly influences the nuclear reaction cross section and imparts special properties to these nuclei \cite{Tanihata,Michael}.

The analysis of proton elastic scattering cross sections for the helium isotopes $^{4}$He, $^{6}$He, and $^{8}$He involves the use of four different phenomenological density distribution models \cite{Alkhazov}. These models---SF (Symmetrized Fermi), GH (Gaussian--Halo), GG (Gaussian--Gaussian), and GO (Gaussian--Oscillator)---are defined on the basis of the point-nucleon density, and each includes two free parameters. These parameters were obtained in Refs.\mbox{ \cite{Alkhazov, Dobrovolsky, Moriguchi}} to reproduce an extended matter density characteristic of helium and lithium isotopes. The SF and GH parameterizations are used to model nucleon distributions in a nucleus without distinguishing between neutrons and protons. These models do not take into account the complex internal structure of many-body nuclear systems. In contrast, the GG and GO parametrizations assume a nuclear composition that includes a core and a halo, where the halo is composed entirely of neutrons, and each component has distinct spatial distributions. This approach provides a more nuanced representation of the nucleus, recognizing the spatial segregation between core nucleons and halo neutrons.

\begin{figure}[H]
\includegraphics[width=1.0\hsize]{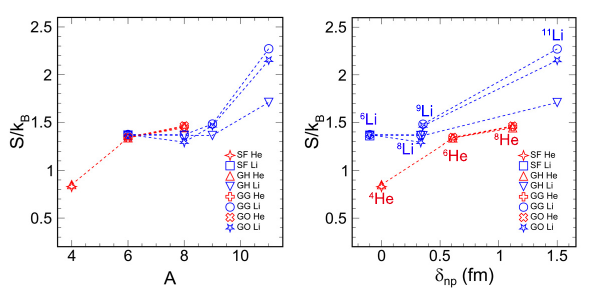}
\caption{({color online} 
) The information entropy for He isotope and Li isotope (\textbf{left}). Correlation of information entropy with neutron skin thickness for He isotope and Li isotope (\textbf{right}).}
\label{fig3}
\end{figure}

Figure \ref{fig3} shows the information entropy for helium and lithium isotopes as determined by four different phenomenological density distribution models. The nucleon distribution in $^{4}\text{He}$ is optimally captured by the SF and GH parameterizations, which treat neutrons and protons equivalently and yield the lowest information entropy. Conversely, the halo nucleus $^{11}\text{Li}$, which is characterized by its core-halo structure, is best described by the GH, GG, and GO models and has the highest information entropy. For $^{6}\text{Li}$, the information entropy values are consistent across all four distribution models. Notably, the neutron-rich helium isotope has a significantly higher information entropy compared to $^{4}\text{He}$. Models GH, GG, and GO are applied to $^{6}\text{He}$, $^{8}\text{He}$, $^{8}\text{Li}$, and $^{9}\text{Li}$. Comparing the information entropy between the neutron-rich nuclei $^{6}\text{He}$, $^{8}\text{He}$, $^{6}\text{Li}$, $^{8}\text{Li}$, and $^{9}\text{Li}$, the values are relatively close.

According to the right side of Figure \ref{fig3}, the information entropy is strongly correlated with the neutron skin thickness. For both lithium and helium isotopes, the information entropy increases as the neutron skin becomes thicker. Although $^{6}\text{He}$ and $^{8}\text{He}$ have larger neutron skin thicknesses, they do not exhibit significantly larger information entropies compared to $^{6}\text{Li}$, $^{8}\text{Li}$, and $^{9}\text{Li}$ due to their similar nucleon numbers. 

Information entropy is a measure of uncertainty and disorder in nuclear density distributions. By evaluating the probability distribution over different points in the nucleus and their associated entropy values, we gain insight into the internal complexity of nuclear structures. Halo nuclei, such as $^{11}$Li, typically have higher entropy because their nuclear density is more radially dispersed. 

\subsection{Application of Information Entropy in Nuclear Cluster Formation}

It has been suggested that atomic nuclei with tetrahedral symmetry could be found throughout the nuclear table \cite{Dudek}. In particular, the tetrahedral structure formed by four $\alpha$ particles in $^{16}$O has been proposed in several studies \cite{Bijker, HeWB, Ding, Ma_NT, Ma_SCP}. The rotation--vibration spectrum of a 4$\alpha$ configuration with tetrahedral symmetry (denoted $T_{d}$) has been studied, and evidence for its occurrence in the lower energy levels of $^{16}$O has been found through experimental excitation spectra. This symmetry-based structure has implications for several areas of physics, suggesting that atomic nuclei with tetrahedral symmetry may be of significant interest.

\begin{figure}[H]
\includegraphics[width=1.0\hsize]{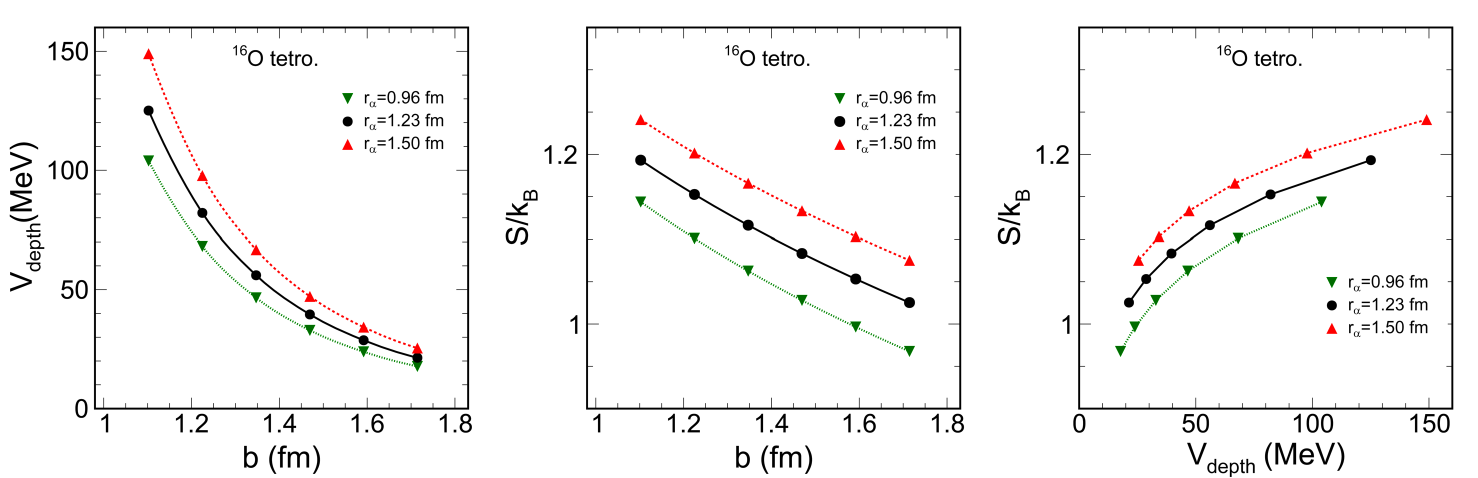}
\caption{({color online} 
) The nuclear potential depth curves for a harmonic oscillator in $^{16}$O with tetrahedral cluster structure (\textbf{left}). The correlation of the information entropy with the cluster dispersion (\textbf{center}). The correlation of the information entropy with the depth of the nuclear potential for a harmonic oscillator (\textbf{right}).}
\label{fig4}
\end{figure}

It was shown that the conditions for cluster formation can be traced back in part to the depth of the confining nuclear potential using the theoretical framework of energy density functionals \cite{Ebran, Ebran1}. The depth of the  nuclear potential for a harmonic oscillator is given by
\begin{equation}
\begin{aligned}
V_{\text{depth}}=\frac{\hbar^{2}}{2m}\frac{R^{2}}{b^{4}},
\end{aligned}
\end{equation}
where $R$ is the radius of the system, $m$ is the mass of the nucleon, $\hbar$ is the reduced Planck constant, and $b$ is the nucleon dispersion in the $\alpha$ cluster distribution. As the dimensionless ratio (of $b$ to the typical interfermion distance $r_{0}$, quantifying nuclear clustering) increases, the nuclear system transitions from a crystalline to a clustered and then to a quantum liquid phase. The trend curves shown on the left of Figure~\ref{fig4} are the depth curves of the nuclear potential versus $b$ for three geometric situations characterized by $r_{\alpha}$, the distance of the center of the $\alpha$-cluster relative to the center of $^{16}$O.

To study the role of information entropy in $^{16}$O with tetrahedral cluster structure, where the distribution of nucleons within an alpha particle based on the alpha particle model is Gaussian type \cite{RBijker},
\begin{equation}
\begin{aligned}
\rho(\vec{r})&=\sum_{i=1}^{4}m_{i}(\frac{\delta_{i}}{\pi})^{3/2}e^{-\delta_{i}(\vec{r}-\vec{r}_{i})^{2}},
\end{aligned}
\end{equation}
where $m_{i}=m_{\alpha}$ is the mass of the $\alpha$ cluster; $\delta_{i}=\delta_{\alpha}=0.52$ is the nucleon distribution deviation for parameterizing the dispersion of the $\alpha$ cluster; $\vec{r}_{i}=\vec{r}_{\alpha}$ is the vector of the $\alpha$ cluster relative to the center of mass of $^{16}$O. The centers of each $\alpha$ cluster are defined as $A=\{\beta,\frac{\pi}{2}+\text{arctan}(\frac{\sqrt{2}}{4}),\frac{7\pi}{6}\}$; $B=\{\beta,\frac{\pi}{2}+\text{arctan}(\frac{\sqrt{2}}{4}),\frac{11\pi}{6}\}$; and $C=\{\beta,\frac{\pi}{2}+\text{arctan}(\frac{\sqrt{2}}{4}),\frac{\pi}{2}\}$; $D=\{\beta,0,0\}$, where $\beta=\frac{\sqrt{6}}{4}\lambda$, and $\lambda$ is the edge length (the center distance of any two $\alpha$ clusters). $\beta=r_{\alpha}$ for the tetrahedral structure of $^{16}$O.

The correlation curves of the information entropy with the cluster dispersion are shown in the middle of Figure~\ref{fig4} for three geometric configurations characterized by the parameter $r_{\alpha}$. The information entropy tends to decrease monotonically as the dispersion parameter $b$ increases. A larger $r_{\alpha}$ leads to a larger system radius, which results in a higher information entropy. The right panel of Figure \ref{fig4} shows some correlation curves between information entropy and potential depth. As the potential depth gradually decreases, the information entropy also decreases.

Studies \cite{Ebran, Ebran1} have concluded that the formation of nuclear clusters can be partially attributed to the depth of the confining nuclear potential, and it has been observed that as the parameter $b$ increases, the nuclear system undergoes a phase transition from a crystalline state to a clustered state and finally to a quantum liquid state. The present investigations show a significant correlation between the information entropy and the parameter $b$. Consequently, it {could be}  reasonable to say that variations in the information entropy are indicative of the nuclear phase transitions.

When fitting the correlation points for a given value of $r_{\alpha}$ (i.e., a given size of the system) through a simple relation
\begin{equation}
\begin{aligned}
S/k_{B}&=\eta\ln \frac{V_{\text{depth}}}{V_{0}},
\end{aligned}
\end{equation}
where $\eta$ and $V_{0}$ are fitting parameters, we can obtain a good prediction curve of $S/k_{b}$ as a function of $V_{\text{depth}}$ with least squares fitting accuracy $\chi^{2}$ of order $10^{-5}$. Equation (7) establishes a relationship between nuclear depth and information entropy, linking the microstate to randomness. A higher entropy indicates a higher degree of disorder. The depth of the nuclear potential determines the degree of confinement and energy level structure of the nucleons in the nucleus, which affects the quality, size, and shape of the nucleus, as well as nuclear reactions. The analysis of Figure~\ref{fig4} suggests that for $^{16}$O, which is characterized by a tetrahedral cluster structure, an increase in the depth of the nuclear potential corresponds to a greater localization of nucleons within the clusters, while higher entropy indicates increased disorder. This suggests an inherent correlation between nuclear potential depth and information entropy.

As the dispersion parameter $a$ increases, the central density of the Woods--Saxon type $^{16}$O relatively decreases, leading to a larger spread of density at a larger radius and, consequently, to an increase in information entropy (Figure~\ref{fig1}). For $^{16}$O with an alpha cluster structure, for a given alpha spacing, increasing the cluster dispersion parameter $b$ causes the central density of the $^{16}$O distribution to increase, making it more like a liquid phase. As a result, the information entropy decreases with increasing $b$ (Figure~\ref{fig4}), which corresponds to the situation of a lower dispersion parameter $a$ for the Woods--Saxon type. In general, this illustrates that the more centrally distributed the nucleons are, the lower the information entropy of the system will be, and vice versa. 

Information entropy is proposed as an available tool for the analysis of nuclear clustering mechanisms. By calculating the information entropy, the degree of disorder in the nuclei with cluster structures can be revealed, and their formation mechanisms can be further understood. Nuclear clustering is an important aspect of nuclear structure, and this clustering behavior is mainly observed in light nuclei, leading to unique nuclear properties that deviate from traditional mean-field models. In this case, a higher entropy value indicates a more dispersed and disordered cluster structure, while a lower entropy value implies a more centralized arrangement, like a quantum liquid phase. 

\section{Summary}
This study investigates the role of information entropy in nuclear physics, focusing on nuclear density distributions, including halo structures and alpha clustering. Information entropy measures uncertainty and disorder, providing insight into the internal complexity of nuclear structures. There is a linear relationship between information entropy and diffusion parameters, indicating increasing uncertainty in the radial nucleon distribution with increasing diffusion. Heavier nuclei typically have higher information entropy, suggesting greater complexity in larger systems. 

In addition, there is a strong correlation between information entropy and neutron skin thickness, which is an important parameter for understanding nuclear symmetry energy. The analysis of helium and lithium isotopes shows different levels of information entropy, with the halo nucleus $^{11}$Li showing the highest entropy due to its widely dispersed core-halo structure. On the other hand, the most tightly bound nucleus, $^{4}$He, has the lowest level of information entropy. 

For $^{16}$O, which is characterized by a tetrahedral cluster structure, the information entropy decreases as the potential depth gradually decreases. By studying the difference in information entropy in a nucleus, we can gain insight into the stability and dynamics of nuclear clusters. It underscores the value of information entropy as a key tool for revealing and exploring the diversity of nuclear structures.

Current models of nuclear density distribution, while phenomenological rather than based on microscopic quantum mechanics, effectively capture the approximate features of nucleon distribution in atomic nuclei. Information entropy offers a novel perspective by quantifying the diversity of nuclear structures, thereby improving our understanding of nuclear systems. However, these models are limited in that they do not fully capture the subtleties of quantum effects at the microscopic level. In future studies, we aim to further validate the current concepts and results by incorporating microscopic density distributions of atomic nuclei.

\vspace{6pt}
\authorcontributions{{Conceptualization, Y.G. Ma and W.H. Ma; methodology, Y.G. Ma and W.H. Ma; software, W.H. Ma; validation, Y.G. Ma and W.H. Ma; formal analysis, Y.G. Ma and W.H. Ma; investigation, Y.G. Ma and W.H. Ma; resources, Y.G. Ma and W.H. Ma; data curation, W.H. Ma; writing---original draft preparation, W.H. Ma; writing---review and editing, Y.G. Ma; visualization, W.H. Ma; supervision, Y.G. Ma; project administration, Y.G. Ma; funding acquisition, Y.G. Ma and W.H. Ma. All authors have read and agreed to the published version of the manuscript.} 
}

\funding{{This work is supported by the Natural Science Foundation of Shanghai with Grants No. 23JC1400200, the National Natural Science Foundation of China with Grants No. 11905036,  12147101 and 11890710, the National Key R\&D Program from the Ministry of Science and Technology of China (2022YFA1604900), and the STCSM under Grant No. 23590780100.} 
}

\dataavailability{{No new data were created or analyzed in this study. Data sharing is not applicable to this article.} 
}

\acknowledgments{{We have no additional acknowledgments to make.}  
}

\conflictsofinterest{The authors declare no conflicts of interest{.} 
}

\begin{adjustwidth}{-\extralength}{0cm}

\reftitle{References}

\PublishersNote{}
\end{adjustwidth}
\end{document}